\documentclass[3p,times,authoryear]{elsarticle}
\usepackage{amsmath,amssymb,amsthm,booktabs}
\usepackage{mathrsfs}
\usepackage{pifont}
\usepackage[OT1]{fontenc}
\usepackage[utf8]{inputenc}
\usepackage{txfonts}
\usepackage{indentfirst}
\usepackage{multirow}
\usepackage{float,subfig}

\usepackage{bm}
\usepackage{euscript}
\usepackage{graphicx}
\usepackage{multicol}
\usepackage[usenames,dvipsnames,svgnames,table]{xcolor}
\usepackage[colorlinks,citecolor=blue,urlcolor=blue]{hyperref}

\biboptions{authoryear}
\numberwithin{equation}{section} \theoremstyle{plain}
\newtheorem{theorem}{Theorem}[section]

\newtheorem{remark}{Remark}[section]

\begin{document}

\newcommand\tabfig[1]{\vskip5mm \centerline{\textsc{Insert #1 around here}}  \vskip5mm}

\begin{frontmatter}


\title{A compensatory model for quantile estimation and application to VaR}

\author[inst1]{Xiaoyuan Tian}
\ead{530581277@qq.com}

\author[inst1]{Shuzhen Yang\corref{cor1}}
\ead{yangsz@sdu.edu.cn}

\cortext[cor1]{Corresponding author.}
\affiliation[inst1]{
  organization={Shandong University-Zhong Tai Securities Institute for Financial Studies, Shandong University},
  addressline={South of Shanda Road No 27},
  city={Jinan},
  postcode={250100},
  state={Shandong Province},
  country={China}
}

\begin{abstract}
Unlike the standard two-step workflow of estimating a time series distribution and extracting quantiles from it, this paper proposes a compensatory model to refine quantile estimates based on an existing fitted distribution. We embed a new penalty term in the model and theoretically characterize its ability to bound realized coverage errors, yielding an adaptive quantile estimator. Backtests on the S\&P 500 and NASDAQ Composite show that the compensatory model substantially reduces unconditional coverage errors across four VaR estimators: all 16 compensatory model forecasts pass the unconditional coverage test, compared with 7 of the 16 corresponding Base forecasts. The conditional-calibration results remain estimator-dependent, indicating that compensatory model is a coverage-correction layer rather than a replacement for conditional-tail modelling.
\end{abstract}


\end{frontmatter}

\section{Introduction}

For a given time series, quantiles are important statistics that capture different distributional characteristics, especially tail behavior. 
Traditional methods first estimate the underlying distribution and then calculate its quantile, 
for example by assuming a normal, empirical, $t$, or filtered distribution. 
Related work includes extreme quantile estimation, nonparametric quantile estimation, quantile forecasting, and quantile regression \citep{HL84,DH89,TL06,G11,K17}.
An important application is Value at Risk (VaR), a common measure of downside risk. Many studies have evaluated VaR models and distribution settings; see, for example, \citet{Kuester06}, \citet{Jorion10}, \citet{ABL14}, and \citet{Zhang17}. The literature includes ARCH and GARCH models \citep{Eng82,Bollerslev86}, conditional autoregressive VaR models \citep{EM04}, time-varying quantile models \citep{RH06,Rh09}, and recent work on VaR and expected shortfall \citep{MT20,WZL2020}. Furthermore, VaR accuracy depends on both the forecasting framework and backtesting specifications \citep{SL20,EHHLSY23}.

In this study, we construct a compensatory model (COMM) to improve the estimation of quantile from a given distribution estimator. Although this estimate may give a reasonable quantile in a conventional sense, the realized frequency count can still deviate from the target level. Let $\{X_n\}_{n=1}^{\infty}$ be a time series and each $X_n$ follows distribution $F_n(\cdot)$, and let $\hat{F}_n(\cdot)$ be the estimate of $F_n(\cdot)$. For quantile level $\alpha$, define $h({X}_n)=1$ when $X_n\leq q_{\alpha}(\hat{F}_n(\cdot))$ and $0$ otherwise. If $\hat{\alpha}(\boldsymbol{X}_n)=\displaystyle \frac{\sum_{i=1}^nh(X_i)}{n}$ converges to $\alpha$, then $q_{\alpha}(\hat{F}_n(\cdot))$ is a "good" quantile estimate, where $\boldsymbol{X}_n=(X_1,\cdots,X_n)$.
This criterion is relevant for VaR, whose quality is typically assessed by the frequency of realized violations.
A fitted distribution may describe the historical sample, but its induced quantile can still generate too many or too few violations. This motivates a correction applied after the distribution estimator has been obtained.

To improve the performance of $q_{\alpha}(\hat{F}_n(\cdot))$, we introduce a penalty term in the definition of the function $h(\boldsymbol{X}_n)$, which equals $1$ if $X_n\leq q_{\alpha}(\hat{F}_n(\cdot))-\kappa_{\alpha}(n)\left[\hat{\alpha}(\boldsymbol{X}_{n-1})-\alpha\right] $, and $0$ otherwise, where $\kappa_{\alpha}(n)>0$ increases to infinity with $n\to \infty$, and denotes the control ability of the penalty term. We call $q_{\alpha}(\hat{F}_n(\cdot))-\kappa_{\alpha}(n)\left[\hat{\alpha}(\boldsymbol{X}_{n-1})-\alpha\right] $ the adjusted quantile estimate of $X_n,\ n\geq 1$. In theory, we prove that the penalty term can control the convergence error of the count function for bounded and unbounded time series. We then consider the application of the adjusted quantile estimation for VaR. The empirical analysis shows that COMM improves unconditional coverage under several Base distribution estimators, although its broader forecast-quality effects depend on the underlying estimator.
This design differs from choosing a more complicated distributional model. The adjustment only uses historical deviation from the target level, so it calibrates a quantile forecast when misspecification causes repeated underestimation or overestimation of violations.
The proposed method is therefore especially suitable for situations in which the researcher wants to keep a simple benchmark distribution, such as a rolling normal distribution, but still improve the realized coverage of the resulting quantile. In this sense, COMM works as a feedback correction rather than as a new parametric specification.

Unlike the adaptive conformal inference frameworks of \cite{GC2021} and \cite{ACT2023}, which are primarily designed for online prediction sets and distribution-free uncertainty quantification, our compensatory model operates directly on quantile estimates derived from a pre-specified parametric or semi-parametric distribution. While their methods rely on general conformal scores and control-theoretic update rules to achieve marginal coverage under arbitrary distribution shifts (\cite{GC2024}), our approach embeds a penalty term tied to the cumulative violation-frequency error, with explicit theoretical bounds on realized coverage deviations for both bounded and unbounded time series. Crucially, compensatory model is model-agnostic and serves as a post-estimation calibration layer rather than a replacement for conditional-tail modeling, preserving the underlying distributional structure while correcting unconditional coverage bias. This distinguishes our work from conformal prediction: we establish almost-sure convergence of the adjusted quantile under fixed distribution estimators, and show via VaR backtesting that the correction reliably removes unconditional coverage bias but does not universally enhance conditional calibration—confirming its role as a feedback-based coverage corrector, not a general-purpose adaptive inference engine.

This study makes three contributions. First, it develops a feedback-based quantile adjustment and characterizes its coverage-control properties for bounded and unbounded time series. Second, COMM is model-agnostic: it operates as a post-estimation calibration layer and can therefore be combined with different distribution estimators without replacing their underlying specifications. Third, the empirical analysis evaluates the method across two equity indices, two tail probabilities, and four Base VaR estimators, thereby documenting both its broad unconditional-coverage gains and the estimator dependence of conditional calibration.

This remainder of this paper is organized as follows: In Section \ref{sec:cm}, count functions are constructed for bounded and unbounded time series. For the bounded time series, we prove that the count function converges to $\alpha$, $P-a.s.$. For the unbounded case, we establish analogous convergence results. Section \ref{sec:ea} evaluates COMM on the S\&P 500 and NASDAQ Composite and reports the feedback-coefficient selection check. Section \ref{sec:con} concludes the study.

\section{Compensatory Model for Quantile Estimation}\label{sec:cm}

For a given sequence $\{X_n\}_{n=1}^{\infty}$ defined on a probability space $(\Omega,\mathcal{F},P)$, each $X_n$ is a random variable, and $\mathbb{E}[\cdot]$ is the expectation under probability $P$. Here, $X_n$ can be used to represent the return rate of risky assets, or a sample of statistical population. In the following, we introduce COMM for the quantile estimation of the bounded and unbounded time series.

\subsection{Bounded time series}

Let the distribution estimator of $X_n$ be $\hat{F}_n(\cdot)$, the true distribution of $X_n$ be $F_n(\cdot)$, $n\geq 1$. For a given $\alpha\in (0,1)$, let $q_{\alpha}(\hat{F}_n(\cdot))$ be the $\alpha$ quantile of the distribution $\hat{F}_n(\cdot)$. Based on the historical information $\{X_s\}_{s<n}$, we can obtain $\hat{F}_n(\cdot)$ and $q_{\alpha}(\hat{F}_n(\cdot))$, which is obtained by minimizing the check function
$$\sum_{X_s<\xi}(\alpha-1)(X_s-\xi)+\sum_{X_s\geq\xi}\alpha(X_s-\xi).$$
If the count function $\displaystyle\frac{\sum_{n=1}^mI(X_n<q_{\alpha}(\hat{F}_n(\cdot)))}{m}$ converges to $\alpha$ as $m\to \infty$, then $q_{\alpha}(\hat{F}_n(\cdot))$ can be considered a good quantile estimate of the sequence $\{X_n\}_{n=1}^{\infty}$. However, this count function can be far away from $\alpha$. Thus, we add a penalty term in the quantile estimation $q_{\alpha}(\hat{F}_n(\cdot))$, and denote by
\begin{equation}\label{adj-1}
q^{\text{adj}}_{\alpha}(\hat{F}_n(\cdot))
=q_{\alpha}(\hat{F}_n(\cdot))-\kappa_{\alpha}(n)\left[\hat{\alpha}(\boldsymbol{X}_{n-1})-\alpha\right], \ n\geq 2,
\end{equation}
where
$$
\hat{\alpha}(\boldsymbol{X}_{n-1})=\displaystyle \frac{\sum_{i=1}^{n-1}h(\boldsymbol{X}_{i})}{n-1},
$$
\begin{equation}\label{def:01}
h(\boldsymbol{X}_n)=\left\{\begin{array}{rl}
1, &  X_n\leq q^{\text{adj}}_{\alpha}(\hat{F}_n(\cdot)) \\
0, & \text{else},
\end{array}\right.
\end{equation}
and $\boldsymbol{X}_n=(X_1,X_2,\cdots,X_n)$, $\kappa_{\alpha}(n)$ is an increasing function of $n$, and $\hat{\alpha}(\boldsymbol{X}_{0})=\alpha$.

Let $\kappa_{\alpha}=0$ in (\ref{adj-1}). Then $q^{\text{adj}}_{\alpha}(\hat{F}_n(\cdot))$ is the usual estimator of the $\alpha$ quantile of $X_n$ under distribution estimator $\hat{F}_n(\cdot)$. In general, $\{X_n\}_{n=1}^{\infty}$ may not be an independent sequence, and the realized frequency count may not converge to $\alpha$. Thus, the penalty term $\kappa_{\alpha}(n)\left[\hat{\alpha}(\boldsymbol{X}_{n-1})-\alpha\right]$ is used to reduce the distance between $\hat{\alpha}(\boldsymbol{X}_{n-1})$ and $\alpha$. When $\hat{\alpha}(\boldsymbol{X}_{n-1})-\alpha>0$, the value of quantile estimation is too high; thus we subtract the term $\kappa_{\alpha}(n)[\hat{\alpha}(\boldsymbol{X}_{n-1})-\alpha]$ in the adjusted quantile. The case $\hat{\alpha}(\boldsymbol{X}_{n-1})-\alpha<0$ is adjusted in the opposite direction.

The main results of this study are the following. 
\begin{theorem}\label{the-main}
Suppose that there exists a positive constant $L$ such that $\left| X_n \right| \leq L$ and $\left|q_{\alpha}(\hat{F}_n(\cdot))\right|\leq L$ for all $n\geq 1$. Let $\frac{2L}{\kappa_{\alpha}(n)}$  converge to $0$ as $n\to\infty$.
Then
\begin{equation}\label{eq:02}
 P(\lim_{n\to \infty}\bigg\{\left| \hat{\alpha}(\boldsymbol{X}_{n})-\alpha\right|  \leq \frac{2L}{\kappa_{\alpha}(n)}+\frac{1}{n}\bigg\})=1.
\end{equation}
Thus, $\hat{\alpha}(\boldsymbol{X}_{n})$ converges to $\alpha$, $P-a.s.$.  
\end{theorem}

\noindent \textbf{Proof}: For a given $\omega\in N$ and $P(N^C)=0$, if there exists an integer $n_1$ such that
$$
 \hat{\alpha}(\boldsymbol{X}_{n_1}(\omega))-\alpha > \frac{2L}{\kappa_{\alpha}(n_1+1)},
$$
and thus
$$
{\kappa_{\alpha}(n_1+1)}[\hat{\alpha}(\boldsymbol{X}_{n_1}(\omega))-\alpha] > {2L},
$$
$$
q_{\alpha}(\hat{F}_{n_1+1}(\cdot))-{\kappa_{\alpha}(n_1+1)}[\hat{\alpha}(\boldsymbol{X}_{n_1}(\omega))-\alpha] <-L.
$$
By the definition of $h(\boldsymbol{X}_{n_1+1})$ in (\ref{def:01}), it follows that,
$$
h(\boldsymbol{X}_{n_1+1}(\omega))=0.
$$
Note that, if
$$
 \hat{\alpha}(\boldsymbol{X}_{n_1+1}(\omega))-\alpha > \frac{2L}{\kappa_{\alpha}(n_1+2)}.
$$
Similar to the above step, we can obtain that  $h(\boldsymbol{X}_{n_1+2}(\omega))=0$. Furthermore, when
$$
 \hat{\alpha}(\boldsymbol{X}_{n}(\omega))-\alpha > \frac{2L}{\kappa_{\alpha}(n+1)},\ n\geq n_1,
$$
we have
$$
\hat{\alpha}(\boldsymbol{X}_{n}(\omega))=\displaystyle \frac{\sum_{i=1}^{n}h(\boldsymbol{X}_{i}(\omega))}{n}
$$
decreases with $n$ and converges to $0$ as $n\to\infty$. Thus, there exists $n_2>n_1$ such that
\begin{equation}\label{eq:03}
\hat{\alpha}(\boldsymbol{X}_{n_2}(\omega))-\alpha \leq \frac{2L}{\kappa_{\alpha}(n_2+1)}.
\end{equation}

Next, we calculate the distance between $\hat{\alpha}(\boldsymbol{X}_{n_2}(\omega))$ and $\hat{\alpha}(\boldsymbol{X}_{n_2+1}(\omega))$
\begin{equation}
\begin{array}{rl}
&\hat{\alpha}(\boldsymbol{X}_{n_2+1}(\omega))-\hat{\alpha}(\boldsymbol{X}_{n_2}(\omega))\\
=&\displaystyle \frac{\sum_{i=1}^{n_2+1}h(\boldsymbol{X}_{i}(\omega))}{n_2+1}
-\frac{\sum_{i=1}^{n_2}h(\boldsymbol{X}_{i}(\omega))}{n_2}\\
=&\displaystyle\frac{h(\boldsymbol{X}_{n_2+1}(\omega))}{n_2+1}-
\frac{\sum_{i=1}^{n_2}h(\boldsymbol{X}_{i}(\omega))}{n_2(n_2+1)}
\end{array}
\end{equation}
From $h(\boldsymbol{X}_{i}(\omega))\in [0,1]$, we have
$$
\left|\hat{\alpha}(\boldsymbol{X}_{n_2+1}(\omega))-\hat{\alpha}(\boldsymbol{X}_{n_2}(\omega))\right|
\leq \frac{1}{n_2+1}.
$$
Combining inequality (\ref{eq:03}), it follows that
$$
\hat{\alpha}(\boldsymbol{X}_{n_2+1}(\omega))-\alpha \leq \frac{2L}{\kappa_{\alpha}(n_2+1)}+\frac{1}{n_2+1}.
$$
Furthermore, if
$$
\frac{2L}{\kappa_{\alpha}(n_2+1)}<\hat{\alpha}(\boldsymbol{X}_{n_2+1}(\omega))-\alpha \leq \frac{2L}{\kappa_{\alpha}(n_2+1)}+\frac{1}{n_2+1},
$$
we can repeat the above process, and we have for $n\geq n_2$,
$$
\hat{\alpha}(\boldsymbol{X}_{n}(\omega))-\alpha \leq \frac{2L}{\kappa_{\alpha}(n)}+\frac{1}{n}.
$$

Similarly, we can prove that there exists $n_3$ such that when $n\geq n_3$,
$$
\hat{\alpha}(\boldsymbol{X}_{n}(\omega))-\alpha \geq -\frac{2L}{\kappa_{\alpha}(n)}-\frac{1}{n}.
$$
Thus, for $n\geq \max(n_2,n_3)$, we have
$$
\left| \hat{\alpha}(\boldsymbol{X}_{n}(\omega))-\alpha\right|  \leq \frac{2L}{\kappa_{\alpha}(n)}+\frac{1}{n},
$$
which completes the proof.
$ \qquad   \Box$

\begin{remark}\label{re-2}
In the proof of Theorem~\ref{the-main}, we establish that
\[
\left|\hat{\alpha}(\boldsymbol{X}_{n+1}(\omega))-\hat{\alpha}(\boldsymbol{X}_{n}(\omega))\right|
\leq \frac{1}{n+1}.
\]
To ensure the almost sure convergence of $\hat{\alpha}(\boldsymbol{X}_{n}(\omega))$ to $\alpha$, 
we assume that $\frac{2L}{\kappa_{\alpha}(n)} \to 0$ as $n \to \infty$. 
In practice, we typically take $\kappa_{\alpha}(n) =\beta\cdot n$, where $\beta$ is a given constant. We take $\beta\in [0.007,0.003]$ for practical analysis.
\end{remark}

\subsection{Unbounded time series}

In practice, applying Theorem \ref{the-main}, we can deal with the unbounded time series $\{X_n\}_{n=1}^{\infty}$. We use data $\{X_n\}_{n=1}^{m}$ to estimate and modify quantile $q_{\alpha}(\hat{F}_n(\cdot))$. For data $\{X_n\}_{n=1}^{m}$, we can find a constant $L(m)$ such that $\left| X_n\right|\leq L(m),\ n\leq m$. Let $\frac{L(m)}{\kappa_{\alpha}(m)}$ converge to $0$ as $m\to\infty$. Then, by Theorem \ref{the-main}, we have 
$$\left| \hat{\alpha}(\boldsymbol{X}_{m})-\alpha\right|  \leq \frac{2L(m)}{\kappa_{\alpha}(m)}+\frac{1}{m}, \ P-a.s..$$

The bounded and unbounded results have the same implication: the compensatory term reduces the deviation between the realized frequency count and the target quantile level. The model can therefore be viewed as a post-estimation calibration step for any chosen distribution estimator.

\section{Empirical Design and Results}\label{sec:ea}

\subsection{Implementation}

We evaluate COMM as a calibration layer applied to four commonly used one-step-ahead VaR estimators. Let $s=1,\ldots,N$ index forecasts after the initialization window, and let $n_s=W+s$ be the corresponding full-sequence index. Let $q^0_{\alpha,s}$ denote the Base quantile forecast, and the compensated forecast is
\begin{equation}\label{eq:empirical-comm}
q^{\mathrm{COMM}}_{\alpha,s}
=q^0_{\alpha,s}-\beta n_s\left(a^{\mathrm{state}}_{s-1}-\alpha\right),
\qquad \beta=0.005,
\end{equation}
where
\begin{equation}\label{eq:empirical-state}
a^{\mathrm{state}}_{s-1}
=\frac{\alpha W+\sum_{i=1}^{s-1}h_i}{W+s-1},
\qquad
h_i=\mathbf{1}\!\left\{X_i\leq q^{\mathrm{COMM}}_{\alpha,i}\right\}.
\end{equation}
The pseudo-count $\alpha W$ sets $a^{\mathrm{state}}_0=\alpha$ and stabilizes the feedback state at the beginning of the recursion. It is not included in the reported violation rate. Because $n_s=W+s$, equations~(\ref{eq:empirical-comm})--(\ref{eq:empirical-state}) implement the theoretical rule $\kappa_\alpha(n)=\beta n$. All reported coverage statistics use the realized count $N_1=\sum_{s=1}^{N}h_s$ and the evaluation-sample rate $N_1/N$.

The Base forecasts comprise a rolling normal estimator, a rolling historical quantile, a GARCH$(1,1)$ model with Student-$t$ innovations, and a filtered historical simulation (FHS). The rolling normal and historical estimators are standard parametric and nonparametric VaR benchmarks, respectively \citep{Kuester06}. The GARCH$(1,1)$ specification follows \citet{Bollerslev86}, with Student-$t$ innovations as in \citet{Bollerslev87}, while FHS follows the volatility-filtered simulation approach of \citet{BGV}. The two rolling estimators use $W=200$ observations. GARCH-$t$ is re-estimated recursively with at most 1,000 preceding returns, whereas FHS takes the empirical tail quantile of standardized residuals and rescales it by the current conditional volatility. We examine $\alpha\in\{0.01,0.05\}$.

Performance is assessed by the unconditional coverage (UC) test of \citet{Kupiec95} and the independence (IND) and conditional coverage (CC) tests of \citet{Christoffersen98}. We also record mean pinball loss, based on the quantile check-loss criterion of \citet{KoenkerBassett78}, and the mean absolute daily change in the VaR forecast. The latter measures path variation rather than forecast accuracy. A test is classified as passed when its $p$-value is at least 0.05. For both Base--COMM comparisons and comparisons among candidate feedback coefficients, uncertainty in paired daily loss differences is evaluated by a moving-block bootstrap \citep{Kunsch89} with a 20-day block length and 5,000 replications.

\subsection{Data and evaluation protocol}

The main application uses daily closing prices of the S\&P 500 downloaded from Yahoo Finance from 20 March 2017 to 29 July 2026. Log returns are calculated from adjacent closing prices, and the retained return series begins on 22 March 2017 according to the prespecified sample cutoff. The first retained value is therefore the return from the close on 21 March to the close on 22 March. The first 200 returns initialize the rolling estimators, and forecasts through 31 December 2019 initialize the recursive state. The reported evaluation sample runs from 2 January 2020 to 29 July 2026 and contains $N=1{,}651$ one-step-ahead forecasts.
The NASDAQ Composite provides a second-market check under the same initialization rule, tail probabilities, and Base estimators. Its evaluation sample contains the same 1,651 trading days as the S\&P 500, with an identical date sequence from 2 January 2020 to 29 July 2026. 

The final design therefore contains two indices, two tail probabilities, and four Base estimators, yielding 16 forecast comparisons. To separate parameter tuning from final evaluation, we select the feedback coefficient using only the common pre-2020 validation period from 5 January 2018 to 31 December 2019, comprising 500 forecasts for each market--tail--estimator combination. Among the candidate values $\beta\in\{0.003,0.004,0.005,0.006,0.007\}$, the coefficient minimizing the equally weighted mean pinball loss across the 16 combinations is retained for the 2020--2026 evaluation.

\subsection{S\&P 500: mechanism and model-level evidence}

\begin{figure}[!htbp]
\centering
\includegraphics[width=0.96\linewidth]{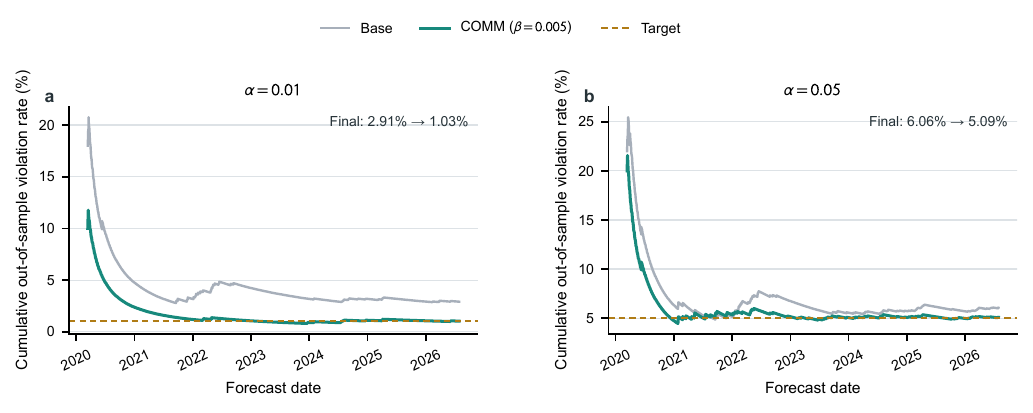}
\caption{Cumulative out-of-sample violation rates for the rolling normal VaR estimator on the S\&P 500. The grey curve is the Base forecast, the green curve is COMM with $\beta=0.005$, and the dashed line is the nominal target. Final rates use the complete evaluation sample.}
\label{fig:sp500-coverage}
\end{figure}

Figure~\ref{fig:sp500-coverage} illustrates the adjustment mechanism for the rolling normal estimator. To avoid visually unstable ratios at the start of the evaluation sample, the paths are displayed after the first 50 forecasts; the final annotations and all formal tests nevertheless use all 1,651 observations. At both tail probabilities $\alpha=0.01$ and $\alpha=0.05$, the Base violation rate remains above its target, whereas the COMM path moves rapidly towards the nominal level and subsequently remains close to it.

Table~\ref{tab:sp500-main} shows that the coverage improvement is not confined to the normal benchmark. At $\alpha=0.01$, the COMM violation rates range from 1.03\% to 1.09\%. At $\alpha=0.05$, all four compensated estimators yield a rate of 5.09\%. COMM passes the UC test in all eight comparisons, whereas the corresponding Base forecasts pass in four. The IND and CC tests each pass in six COMM comparisons. The remaining IND and CC rejections occur for the normal and historical estimators at $\alpha=0.01$. Pinball loss is lower in five of the eight comparisons, with an average reduction of 2.19\%. No reduction is significant under the paired moving-block bootstrap, whereas the increase for FHS at $\alpha=0.05$ is significant. COMM also increases mean absolute daily VaR variation by 0.024 percentage points on average. Thus, the adjustment consistently corrects unconditional frequency bias, although its broader forecast-quality effects remain estimator-dependent.

\begin{table}[H]
\centering
\caption{S\&P 500 out-of-sample results, 2 January 2020--29 July 2026.}
\label{tab:sp500-main}
\small
\resizebox{\textwidth}{!}{%
\begin{tabular}{llrrrrrr}
\toprule
$\alpha$ & Estimator & Rate (\%, B/M) & UC $p$ (B/M) & IND $p$ (B/M) & CC $p$ (B/M) & $\Delta L$ (\%) & $\Delta S$ (pp) \\
\midrule
0.01 & Normal     & 2.91/1.03 & $<0.01$/0.90 & 0.01/0.01 & $<0.01$/0.04 & $-12.96$ & $+0.01$ \\
0.01 & Historical & 1.94/1.09 & $<0.01$/0.72 & $<0.01$/0.01 & $<0.01$/$<0.05$ & $-2.65$ & $+0.01$ \\
0.01 & GARCH-$t$  & 1.70/1.09 & 0.01/0.72 & 0.50/0.19 & 0.03/0.40 & $-0.71$ & $+0.01$ \\
0.01 & FHS        & 1.21/1.09 & 0.40/0.72 & 0.24/0.19 & 0.35/0.40 & $+1.64$ & $+0.01$ \\
0.05 & Normal     & 6.06/5.09 & 0.06/0.87 & 0.02/0.20 & 0.01/0.43 & $-5.10$ & $+0.04$ \\
0.05 & Historical & 5.57/5.09 & 0.29/0.87 & 0.01/0.19 & 0.03/0.42 & $-4.40$ & $+0.05$ \\
0.05 & GARCH-$t$  & 7.81/5.09 & $<0.01$/0.87 & 1.00/0.91 & $<0.01$/0.98 & $+2.19$ & $+0.04$ \\
0.05 & FHS        & 5.51/5.09 & 0.35/0.87 & 0.99/0.91 & 0.64/0.98 & $+4.46$ & $+0.04$ \\
\bottomrule
\end{tabular}%
}
\begin{minipage}{0.98\textwidth}
\footnotesize \emph{Notes}: B/M denotes Base and COMM, respectively. All numerical entries are reported to two decimal places. Rate is the realized violation rate in percent. $\Delta L=100(L_{\mathrm{COMM}}/L_{\mathrm{Base}}-1)$, so negative values favour COMM. $\Delta S=S_{\mathrm{COMM}}-S_{\mathrm{Base}}$, where $S$ is the mean absolute daily VaR change in percentage points; positive values indicate greater path variation. Small $p$-values are reported by thresholds where rounding could obscure rejection.
\end{minipage}
\end{table}

\subsection{NASDAQ: external-market evidence}

Table~\ref{tab:nasdaq-main} reports the corresponding NASDAQ results over exactly the same evaluation dates as the S\&P 500. At $\alpha=0.01$, COMM produces violation rates between 1.03\% and 1.15\%; at $\alpha=0.05$, the rates range from 4.97\% to 5.09\%. All eight COMM forecasts pass the UC test, compared with three Base forecasts. IND and CC each pass in seven COMM comparisons, compared with five and three Base comparisons, respectively. The only COMM rejection for both IND and CC occurs for the normal estimator at $\alpha=0.01$. Pinball loss is lower in five of eight comparisons and declines by 2.47\% on average. The individual reductions are not significant, while the FHS increase at $\alpha=0.05$ is significant. Mean absolute daily VaR variation increases by 0.024 percentage points on average.

\begin{table}[H]
\centering
\caption{NASDAQ Composite out-of-sample results, 2 January 2020--29 July 2026.}
\label{tab:nasdaq-main}
\small
\resizebox{\textwidth}{!}{%
\begin{tabular}{llrrrrrr}
\toprule
$\alpha$ & Estimator & Rate (\%, B/M) & UC $p$ (B/M) & IND $p$ (B/M) & CC $p$ (B/M) & $\Delta L$ (\%) & $\Delta S$ (pp) \\
\midrule
0.01 & Normal     & 2.91/1.15 & $<0.01$/0.55 & 0.01/$<0.01$ & $<0.01$/$<0.01$ & $-9.79$ & $+0.01$ \\
0.01 & Historical & 2.00/1.09 & $<0.01$/0.72 & 0.03/0.19 & $<0.01$/0.40 & $-2.56$ & $+0.01$ \\
0.01 & GARCH-$t$  & 2.00/1.03 & $<0.01$/0.90 & 0.69/0.55 & $<0.01$/0.83 & $-3.37$ & $+0.01$ \\
0.01 & FHS        & 1.03/1.09 & 0.90/0.72 & 0.17/0.53 & 0.38/0.77 & $+0.75$ & $+0.01$ \\
0.05 & Normal     & 6.72/5.09 & $<0.01$/0.87 & 0.02/0.40 & $<0.01$/0.70 & $-6.28$ & $+0.04$ \\
0.05 & Historical & 6.00/5.09 & 0.07/0.87 & 0.10/0.40 & 0.05/0.70 & $-3.90$ & $+0.05$ \\
0.05 & GARCH-$t$  & 7.45/5.03 & $<0.01$/0.96 & 0.95/0.53 & $<0.01$/0.82 & $+1.44$ & $+0.04$ \\
0.05 & FHS        & 5.57/4.97 & 0.29/0.95 & 0.28/0.64 & 0.33/0.89 & $+4.00$ & $+0.03$ \\
\bottomrule
\end{tabular}%
}
\begin{minipage}{0.98\textwidth}
\footnotesize \emph{Notes}: notation follows Table~\ref{tab:sp500-main}.
\end{minipage}
\end{table}

Table~\ref{tab:beta-selection} reports the validation experiment. All five candidate coefficients perform similarly: each passes the UC test in all 16 combinations, and each passes the CC test in at least 15. Their aggregate mean pinball losses lie within 0.94\% of the minimum. The prespecified loss criterion selects $\beta=0.005$, and all 16 leave-one-combination-out checks retain this value. The paired moving-block bootstrap also identifies no combination in which another candidate has significantly lower validation loss than $\beta=0.005$. Although $\beta=0.007$ gives the smallest coverage error, it produces greater loss and path variation. We therefore retain $\beta=0.005$ as a stable common choice, while recognizing that the neighbouring candidates also provide broadly comparable calibration.

\begin{table}[H]
\centering
\caption{Pre-2020 validation results for feedback-coefficient selection.}
\label{tab:beta-selection}
\small
\resizebox{\textwidth}{!}{%
\begin{tabular}{rrrrrr}
\toprule
$\beta$ & CE (pp) & Mean pinball loss ($\times10^{-3}$) & IND pass & CC pass & Mean $|\Delta\mathrm{VaR}|$ \\
\midrule
0.003 & 0.2000 & 0.886915 & 12/16 & 15/16 & 0.001181 \\
0.004 & 0.1500 & 0.884094 & 13/16 & 16/16 & 0.001229 \\
\textbf{0.005} & 0.0875 & \textbf{0.878657} & 11/16 & 16/16 & 0.001277 \\
0.006 & 0.0875 & 0.881800 & 12/16 & 16/16 & 0.001327 \\
0.007 & \textbf{0.0625} & 0.885264 & 11/16 & 16/16 & 0.001376 \\
\bottomrule
\end{tabular}%
}
\begin{minipage}{0.98\textwidth}
\footnotesize \emph{Notes}: Results aggregate 16 market--tail--estimator combinations over the common validation period from 5 January 2018 to 31 December 2019, using 500 forecasts for each combination. The 2020--2026 evaluation sample is not used to select $\beta$. CE is the mean absolute coverage error in percentage points. IND and CC passes use $p\geq0.05$. Mean $|\Delta\mathrm{VaR}|$ is the mean absolute daily change in the VaR forecast. UC is omitted because every candidate passes all 16 UC tests.
\end{minipage}
\end{table}

\section{Conclusion}\label{sec:con}

This study has developed a compensatory model (COMM) to improve quantile calibration under a given sequence of distribution estimators $\{\hat{F}_n(\cdot)\}_{n=1}^{\infty}$. By adding a feedback penalty to the count function, COMM adjusts each new quantile according to the accumulated violation-frequency error. The theoretical results characterize this coverage-control mechanism. Evidence from the S\&P 500 and NASDAQ Composite shows that COMM with $\beta=0.005$ sharply reduces unconditional coverage errors across four Base VaR families.

The overall forecast-quality improvements depend on the choice of Base estimator. Across the 16 test comparisons, COMM reduces pinball loss in 10 cases and lowers the unweighted mean loss by 2.33\%. The adjustment also increases mean absolute daily VaR variation by 0.024 percentage points on average. Gains are most pronounced when the unadjusted forecast exhibits a persistent coverage bias; in contrast, when an FHS forecast is already close to its nominal coverage rate, compensation may increase pinball loss even as coverage improves. COMM should therefore be interpreted as a coverage-calibration layer rather than a universal replacement for conditional-tail modelling.

\section*{CRediT authorship contribution statement}

First author: Writing - original draft, Data curation, Formal analysis, Writing - review \& editing. Second author: Conceptualization, Methodology, Formal analysis, Writing - original draft, Writing - review \& editing.

\section*{Declaration of competing interest}

The authors report there are no competing interests to declare.

\section*{Data availability}

Data were derived from public-domain resources. Daily index data for the S\&P 500 and NASDAQ Composite used in the empirical analysis were downloaded from Yahoo Finance: \url{https://finance.yahoo.com/lookup}.

\section*{Acknowledgements}

\textbf{Funding:} This work was supported by the National Natural Science Foundation of China (Grant No.12471450) and Taishan Scholar Talent Project Youth Project.

\section*{Declaration of generative AI and AI-assisted technologies}
During the preparation of this work, the authors used Doubao AI for academic writing polishing, sentence revision and typesetting suggestions. The authors reviewed and edited the output as needed and take full responsibility for the content of the published article.

\vspace{-\baselineskip}
\begingroup
\linespread{1.0}\selectfont
\setlength{\bibsep}{0pt plus 0.2ex}
\bibliography{gexp1}

@article{GC2021,
 author = {Gibbs, I. and Candès, E.},
 title = {{Adaptive} Conformal Inference Under Distribution Shift},
 journal = {Advances in Neural Information Processing Systems},
 volume = {34},
 year = {2021}
}

@article{ACT2023,
 author = {Angelopoulos, A. and Candès, E. and Tibshirani, R.},
 title = {{Conformal PID} Control for Time Series Prediction},
 journal = {Advances in Neural Information Processing Systems},
 volume = {36},
 year = {2023}
}

@article{GC2024,
author = {Gibbs, I. and Candès, E.},
title = { {Conformal} inference for online prediction with arbitrary `
distribution shifts},
journal = {Journal of Machine Learning Research},
volume = {25(162)},
pages = {1-36},
year = {2024},
}

@article{G11,
author = {Gneiting, T.},
title = {{Quantiles as optimal point forecasts}},
journal = {International Journal of Forecasting},
volume = {27(2)},
pages = {197--207},
year = {2011},
}

@article{K17,
author = {Korobilis, D.},
title = {{Quantile regression forecasts of inflation under model uncertainty}},
journal = {International Journal of Forecasting},
volume = {33(1)},
pages = {11-20},
year = {2017},
}

@article{MT20,
author = {Meng, X. and Taylor, J. W.},
title = {{Estimating Value-at-Risk and Expected Shortfall using the intraday low and
 range data}},
journal = {European Journal of Operational Research},
volume = {280},
pages = {191--202},
year = {2020},
}

@article{EM04,
author = {Engle, Robert F.  and Manganelli, S. },
title = {{CAViaR}: {Conditional} Autoregressive Value at
Risk by Regression Quantiles},
journal = {Journal of Business \& Economic Statistics},
volume = {22:4},
pages = {367--381},
year = {2004},
}

@article{ABL14,
author = {Abad, P. and Benito, S. and  L\'{o}pez, C. },
title = {A comprehensive review of {Value at Risk} methodologies},
journal = {The Spanish Review of Financial Economics},
volume = {12:1},
pages = {15--32},
year = {2014},
}

@article{RH09,
author = {De Rossi, G. and Harvey, A.C. },
title = {Quantiles, expectiles and splines},
journal = {Journal of Econometrics},
volume = {152},
pages = {179--185},
year = {2009},
}

@article{RH06,
author = {De Rossi, G. and Harvey, A.C. },
title = {Time-varying quantiles},
journal = {Faculty of Economics},
volume = {Cambridge},
pages = {CWPE 0649},
year = {2006},
}

@Article{WZL2020,
  author = 		 {Wang, G. and Zhou, K. and Li, G. and Li, W. K.},
  title = 		 {{H}ybrid quantile estimation for asymmetric power {GARCH} models},
  journal = 	 {Journal of Econometrics},
  year = 		 {2020},
  volume = 	 {https://doi.org/10.1016/j.jeconom.2020.05.005},
  pages = 	 {1-22},
}

@Article{BGV,
  author = 		 {Barone-Adesi, G. and  Giannopoulos, K. and  Vosper, L.},
  title = 		 {{VaR} without correlations for nonlinear portfolios},
  journal = 	 {Journal of Futures Markets},
  year = 		 {1999},
  volume = 	 {19},
  pages = 	 {583-602},
}

@article{Kuester06,
author = {Kuester, K. and Mittnik, S. and Paolella, M. S.},
title = {\text{Value-at-Risk Prediction}: A Comparison of Alternative Strategies},
journal = {Journal of Financial Econometrics},
volume = {4},
number = {1},
pages = {53-89},
year = {2006},
}

@Article{Bollerslev86,
  author = 		 {Bollerslev, T.},
  title = 		 {Generalized Autoregressive Conditional Heteroskedasticity},
  journal = 	 {Journal of Econometrics},
  year = 		 {1986},
  volume = 	 {31},
  pages = 	 {307-327},
}

@Article{Eng82,
  author = 		 {Engle, Robert F.},
  title = 		 {{Autoregressive Conditional Heteroskedasticity with
  Estimates of the Variance of United Kingdom Inflation}},
  journal = 	 {Econometrica},
  year = 		 {1982},
  volume = 	 {50(4)},
  pages = 	 {987-1007},
}

@article{Zhang17,
author = {Y. Zhang and S. Nadarajah},
title = {A review of backtesting for value at risk},
journal = {Communications in Statistics - Theory and Methods},
pages = {1-24},
year  = {2017},
}

@article{Jorion10,
author = {Jorion, Ph.},
title = {Risk Management},
journal = {Annual Review of Financial Economics},
volume = {2},
number = {1},
pages = {347-365},
year = {2010},
}

@article{TL06,
author = {Takeuchi, I. and Le, Q. V. and Sears, T. D. and  Smola, A. J. },
title = {{N}onparametric Quantile Estimation},
journal = {Journal of Machine Learning Research},
volume = {7},
number = {45},
pages = {1231--1264},
year = {2006},
}

@article{DH89,
author = {Dekkers, A. L. M.  and De Hana, L.},
title = {{O}n the Estimation of the Extreme-Value Index and Large Quantile Estimation},
journal = {The Annals of Statistics},
volume = {17},
number = {4},
pages = {1795--1832},
year = {1989},
}

@article{HL84,
author = {Heidelberger, P. and  Lewis, P. A. W. },
title = {{Q}uantile Estimation in Dependent Sequences},
journal = {Operations Research},
volume = {32},
number = {1},
pages = {185--209},
year = {1984},
}

@article{SL20,
author = {Sobreira, N. and Louro, I.},
title = {Evaluation of volatility models for forecasting {Value-at-Risk} and {Expected Shortfall} in the Portuguese stock market},
journal = {Finance Research Letters},
volume = {32},
pages = {101098},
year = {2020},
doi = {10.1016/j.frl.2019.01.010},
}

@article{EHHLSY23,
author = {Ewald, C. O. and Hadina, J. and Haugom, E. and Lien, G. and St{\o}rdal, St{\aa}le and Yahya, M.},
title = {Sample frequency robustness and accuracy in forecasting {Value-at-Risk} for Brent Crude Oil futures},
journal = {Finance Research Letters},
volume = {58},
pages = {103949},
year = {2023},
doi = {10.1016/j.frl.2023.103949},
}

@article{Bollerslev87,
author = {Bollerslev, T.},
title = {A Conditionally Heteroskedastic Time Series Model for Speculative Prices and Rates of Return},
journal = {The Review of Economics and Statistics},
volume = {69},
number = {3},
pages = {542--547},
year = {1987},
doi = {10.2307/1925546},
}

@article{Kupiec95,
author = {Kupiec, Paul H.},
title = {Techniques for Verifying the Accuracy of Risk Measurement Models},
journal = {The Journal of Derivatives},
volume = {3},
number = {2},
pages = {73--84},
year = {1995},
doi = {10.3905/jod.1995.407942},
}

@article{Christoffersen98,
author = {Christoffersen, Peter F.},
title = {Evaluating Interval Forecasts},
journal = {International Economic Review},
volume = {39},
number = {4},
pages = {841--862},
year = {1998},
doi = {10.2307/2527341},
}

@article{KoenkerBassett78,
author = {Koenker, Roger and Bassett, Jr., Gilbert},
title = {Regression Quantiles},
journal = {Econometrica},
volume = {46},
number = {1},
pages = {33--50},
year = {1978},
doi = {10.2307/1913643},
}

@article{Kunsch89,
author = {K{\"u}nsch, Hans R.},
title = {The Jackknife and the Bootstrap for General Stationary Observations},
journal = {The Annals of Statistics},
volume = {17},
number = {3},
pages = {1217--1241},
year = {1989},
doi = {10.1214/aos/1176347265},
}
\endgroup
\end{document}